# The Power of Reciprocal Knowledge Sharing Relationships for Startup Success

Allen, T. J., Gloor, P. A., Fronzetti Colladon, A., Woerner, S. L. & Raz, O.

# The Power of Reciprocal Knowledge Sharing Relationships for Startup Success

Allen, T. J., Gloor, P. A., Fronzetti Colladon, A., Woerner, S. L., & Raz, O.

**Abstract**

**Purpose** – This paper examines the innovative capabilities of biotech start-ups in relation to geographic proximity and knowledge sharing interaction in the R&D network of a major high-tech cluster.

**Design/methodology/approach** – This study compares longitudinal informal communication networks of researchers at biotech start-ups with company patent applications in subsequent years. For a year, senior R&D staff members from over 70 biotech firms located in the Boston Biotech cluster were polled and communication information about interaction with peers, universities, and big pharmaceutical companies was collected, as well as their geo-location tags.

**Findings** – Location influences the amount of communication between firms, but not their innovation success. Rather, what matters is communication intensity and recollection by others. In particular, there is evidence that rotating leadership - changing between a more active and passive communication style – is a predictor of innovative performance.

**Practical implications –** Expensive real-estate investments can be replaced by maintaining social ties. A more dynamic communication style and more diverse social ties are beneficial to innovation.

**Originality/value** – Compared to earlier work that has shown a connection between location, network, and firm performance, this paper offers a more differentiated view; including a novel measure of communication style, using a unique dataset, and providing new insights for firms who want to shape their communication patterns to improve innovation, independently of their location.

# 1. Introduction

In their study of the Boston biotech networks Owen-Smith and Powell (2004) contend that geographic proximity will support innovation, as it leads to knowledge spillover effects benefitting companies co-located in the same cluster. Repeating their study in the same Boston biotech cluster ten years later, we find the opposite: it's not location, that matters, but quality of the network interaction. While Owen-Smith and Powell constructed their network through formal alliances extracted from the Bioscan database, we built our network by manually polling R&D management about their interactions with their peers at other biotech startups in the Boston biotech cluster. We find that while location matters for the quantity of communication, the embeddedness of a startup's R&D members in the core of their peer communication network matters for the innovative capability of their company. Alternating changes in leadership among communication partners is also important for facilitating innovation.

Research up to now has been inconclusive about what matters more for start-up business success: location, or communication. As has been shown repeatedly, geographical proximity cannot be analyzed in isolation; rather proximity should always be examined in combination with other dimensions (Boschma 2003). Many arguments have been made qualitatively about the effectiveness of high-tech businesses clustering in geographic proximity, with the most prominent example being Silicon Valley in California (Saxenian, 2006). While evidence has been found for increased communication between start-ups co-located geographically (Gloor et al., 2008; Allen et al., 2009), the triadic link between increased communication and business success (Raz and Gloor, 2007) on the one hand, and co-location and business success on the other hand (Porter et al., 2005) has been little studied. This research tries to close that gap, by comparing the location of biotech start-ups with the amount and quality of interaction between them, and contrasting both with their innovative capabilities (measured by number of patent applications) (Figure 1).

**Figure 1.** Research Framework.

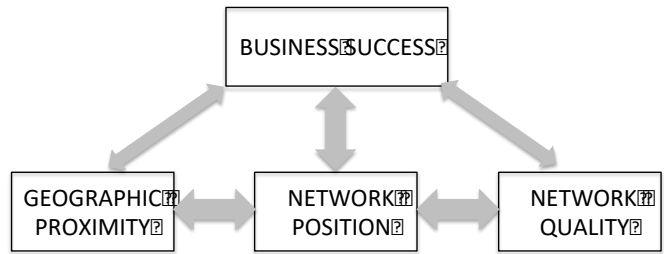

To research this question we drew on a dataset of interpersonal communication of senior R&D members in over 70 firms in the Boston Biotech cluster, with information about firm innovation added. The Boston biotech cluster offers a unique testbed to study communication: firstly the firms in the cluster all belong to the same industry, thus there is no need to control for industry differences, secondly the cluster is big enough to identify and trace communication patterns and thirdly the cluster is host to a number of forums that facilitate inter-firm communication. We focused our analysis on firms working on human therapeutic applications of biologically-derived pharmaceuticals, eliminating companies that are in the agricultural, veterinary, and environmental products and services fields. This left us with about 40 companies inside the Cambridge-Boston cluster. We included about 30 companies within the range of Boston, and added firms from outside the Cambridge-Boston cluster (measured through postal codes) to get a geographical control group.

## 2. Networks as Conduits for Dialog

A large body of literature addresses the topic of network structure, position and performance, with most showing that centrality of individual actors is an important predictor of success. For example, studies show, that in teams of university students, better performing students are more central in network position, (Battistoni and Fronzetti Colladon, 2014), in teams of knowledge workers in multinational corporations, more central team managers lead to more productive teams (Sparrowe et al., 2001; Cummings, 2007), and more communication between Israeli software start-ups during the Internet bubble (Raz and Gloor, 2007) and among fund managers in Canada (Zaheer and Bell, 2005) leads to better fund performance.

Geographic proximity is important to the formation of social ties and amount of communication. In an early study (Festinger et al., 1950) physical proximity predicted the formation of social ties. Allen (1984) demonstrated a strong negative correlation between physical distance and frequency of communication within a firm, and in a later study (Allen et al., 2009) showed that companies in geographic clusters also communicate more. Saxenian (1994) in her study of the electronics industries in Silicon Valley and Boston's Route 128 suggested that dense communication networks are important for innovation and firm success. Geographical proximity among nanotechnology firms is associated with success (Funk, 2013).

The amount of communication between people or between companies seems to be associated with success. Start-ups whose CEOs communicate more with their peers are more successful (Raz and Gloor, 2007). Much key communication in business happens in an informal way, for instance, bumping into a colleague on the way to work, or in a scientific talk at a university – this is the proverbial power of weak ties (Granovetter, 1973).

**Location and Innovation**

A direct relationship between geographic proximity and firm innovativeness, however, has not been as decisively demonstrated. For instance, mutual fund firms within a geographic cluster have been shown to be more innovative than firms on the periphery (Zaheer and Bell, 2005); however the measures of innovation in this study were self-report measures. Owen-Smith and Powell (2004) identified a link between formal alliances among Boston biotech firms, and their membership in the cluster. However, they are not including precise geographical location, and intensity and quality of interaction into their analysis. While they find that being a part of the cluster and being more central increases innovativeness, we speculate that what matters more is quality of the interaction.

This means that just being close to the geographic center of the cluster will increase communication (Allen et al., 2009), but will not increase innovative capabilities of the biotech start-ups. Being embedded in a social network is more important than geographic proximity, as being embedded in the communication network increases business success (Uzzi, 1996; Raz and Gloor, 2007) and helps increase innovation (Uzzi and Spiro, 2005).

Past research has shown that a firm's embeddedness in an alliance network might be a predictor of its innovative performance. However, there can also be too much embeddedness. Studying alliance networks among apparel manufacturers, Uzzi (1996) found an inverse u-shaped relationship between a firm's performance and its embeddedness. Having too many strong ties with business partners can restrict a firm's flexibility in times of crisis. A similar effect was found for academics. In a longitudinal study of an Italian research community, (Rotolo and Petruzzelli 2013) found that more central academics were more productive – however, again up to a point, demonstrating the inverse u-shaped relationship. More ties bridging diverse research communities were also predictors of positive academic performance. Capaldo and Petruzzelli (2014) showed that geographical distance among collaborating business partners had a negative influence on the quality of the collaboration. However, they found that geographic distance and organizational proximity are contigent upon one another in positively influencing innovative performance. We therefore hypothesize a similar relationship between geographical distance and social capital.

*H1a. Firms in the communication core of the cluster are more innovative than those that are in the periphery.*

*H1b. There is no significant effect on innovative capabilities of firms for being close to the geographical center.*

**Network Position and Innovation**

It has repeatedly shown that a preferred position in the network of startups provides preferential access to knowledge, thus assisting in innovation (Uzzi & Spiro, 2005; Cross and Cummings, 2004). We argue that what really matters in creating new knowledge is the quality of the interaction (Goldman and Scardamalia, 2013). Grippa and Gloor (2009) found that remembering social interactions with other members of a research group predicted later success of that research group. Being remembered is often indicative of importance (i.e., socially popular, having access to resources, providing knowledge) and these people are chosen more as discussion partners. We therefore hypothesize that being recalled as a communications partner is a positive predictor of a company's innovative prowess.

*H2. Firms that are mentioned more as communication partners are more innovative.*

**Network Quality and Innovation – Rotating Leadership**

The next question that arises is if there are communication characteristics conducive to an interaction being recalled more frequently. The liveliness of a discussion is one of the characteristics that make a discourse between two communication partners worth remembering. A metric of liveliness is turn-taking, i.e., the number of turns between two discussion participants. Turn-taking in groups has been associated with collective intelligence of the group (Wooley et al., 2010). On the content level, Chuy et al. (2013) argue that taking turns engaging in a knowledge building dialog leads to individual advancement of knowledge. Turn-taking is also associated with transformational leaders, a style of leadership that can lead to more creative individuals and more innovative firms (Gumusluoglu and Ilsev, 2009). Transformational leaders empower their team members by delegating responsibility and encouraging self-organization, fostering intrinsic motivation. This leadership style is reflected in rapidly changing networking structures, where actors fluctuate between peripheral and more central network positions. For instance, in software developer teams, (Kidane and Gloor, 2007) found that a high number of changes in network structure in teams, oscillating between low and high betweenness centrality group structures (Freemann, 1977), correlated with group creativity. Fliaster and Schloderer (2010) show that the amount of knowledge that two actors exchange, the level of responsiveness among them, and the efficiency of knowledge-related collaboration are predictors of creative performance and effectiveness. Davis and Eisenhardt (2011) found that rotating leadership in joint projects between companies is a predictor for the success of the project. In sum, it is not enough for collaborative innovators to just work together, rather they need to frequently exchange the baton.

*H3. Firms that change their network positions more  - from being central to being peripheral, and vice versa – are more innovative.*

## 3. Data and Methods

To test our hypothesis, a unique dataset was collected. For the duration of one year in 2005 and 2006, on a random day once per week, approximately 10% of the R&D staff members of 200 Boston biotech firms were contacted by e-mail, and asked to report with whom they had spoken on that day. Due to data quality issues we ended up with a representative sample of 70 firms. Previous work by Allen et al. (2009) using the same dataset documented communication patterns in the Boston Biotech Cluster, assisted by the Massachusetts Biotechnology Council and the MIT Entrepreneurship Center. Our research builds on this work.

The data set includes only therapeutics firms in the Boston biotech cluster, eliminating those companies primarily focused on agricultural, veterinary and environmental products. Boston is one of the world's largest biotech hubs. According to the Massachusetts Biotechnology Council report (2014), there are more than 900 bio-pharmaceutical and medical technology firms in the Boston area, more than 90 venture capital firms investing in life sciences and 6 out of the top 10 NIH-funded hospitals. The cluster houses the R&D centers of 8 of the top 10 pharmaceutical companies and, collectively, the industry employs more than 56,000 people in the Boston area. In addition, several major universities – including Harvard and MIT – maintain strong relationships with firms in the area.

Focusing on the Boston Biotech Cluster has several advantages: data is homogeneous with respect to the industry; the cluster is big enough to trace and study a significant amount of communication relationships; being in geographical proximity favors firms' communication, alliances and access to knowledge and knowledge workers (Allen et al., 2009; Pouder and St. John, 1996) — the biotech industry is very knowledge intensive. Moreover, the knowledge needed by the biotech firms is highly complex and specific, representing a confluence of multiple disciplines, including chemistry, biology and computer science. Most start-up biotech companies, with limited resources, cannot rely solely on their internal capabilities and intangible assets (Teece, 1986) so we expect that being embedded in a dense communication cluster will favor the capability to generate innovation.

Among the actors in the sample of Allen and colleagues (2009) there are 6 major universities (Boston University, Harvard, MIT, Northeastern, Tufts and UMass Boston), 67 biotech firms and 5 big pharmaceutical companies. For the regression analysis, we included self-reported interactions with five additional types of organizations: the hospitals in the Boston area; the big pharmaceutical companies outside the Boston area; the biotech start-ups outside the Boston area; the universities outside and non-profit research organizations.

*Communication network*

A total of 410 randomly selected scientists from the selected firms were contacted regularly – approximately once a week – during a six-month sliding window within the period 2005-2006.

Each company was represented by approximately 10% of their research personnel (active researchers with PhD or MD degree). Briefly, each participating scientist, each week, received an email message with a link to a specifically developed webpage (Figure 2); on this page they simply had to indicate which company(s) within the Biotech Cluster, outside peer company, university, hospital or large pharmaceutical company(s) they had been in contact with on the day they were answering the survey.

**Figure 2.** Survey screen of data collection system.

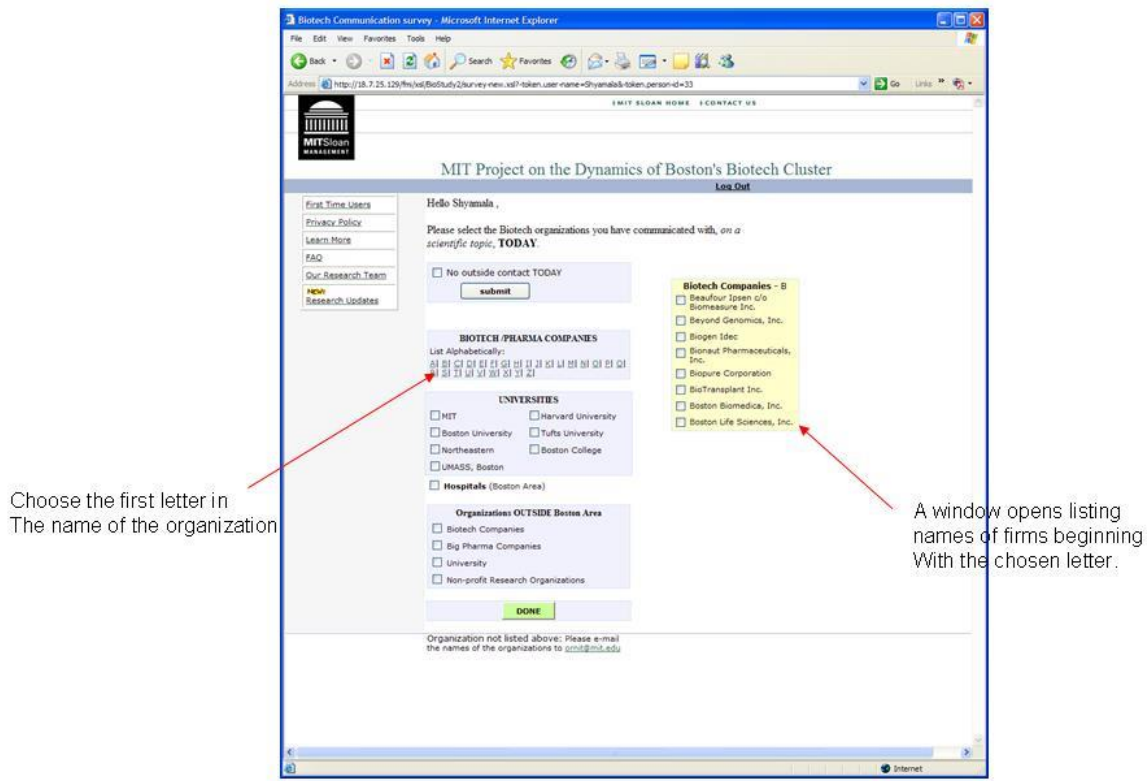

From the collected data, we constructed a communication network, where each firm (or institution) is linked to another, if their R&D departments, represented by the researchers responding to our survey, were communicating during the collection time. We also included a dummy node to keep track of all the communication instances each time the respondents did not specify the peer. Visual and mathematical analysis of the interaction patterns was carried out using the software *Condor*[1](Gloor and Zhao, 2004).

*Geographic proximity*

Geographic proximity was measured by the address of a firm's headquarter in 2005. The distance (expressed in km) has been calculated, for each pair of firms, according to the Haversine formula (Sinnott, 1984), using latitude and longitude coordinates. In addition we assigned each company to either the Cambridge-Boston central cluster, or to the outside.

*Coreness*

Coreness represents the position of the company in the social network of the sample. The core/periphery structure of networks has been frequently addressed in prior research (e.g. Cattani and Ferriani, 2008; Cross and Cummings, 2004; Barsky, 1999). In particular, firms in the core represent a cohesive subgroup and firms in the periphery are loosely connected to the core. We use the approach described by Borgatti and Everett (1999), partitioning the graph into core and periphery, i.e. each firm gets a continuous coreness value and is also assigned either to the periphery or the core.

*Popularity and Creativity*

We model popularity of a startup as its indegree, i.e. the number of companies that reported a startup at least once as a communication partner. In prior work (Kidane and Gloor, 2007) it was found that a change in network position of an actor from core to periphery and vice versa calculated as oscillation in betweenness centrality, i.e. local maxima in the betweenness curve over time (Gloor et al., 2014) is a proxy for its creativity. Formally, we count the local maxima of function $f(t)$=betweenness centrality$(t)$ within time interval $[t_1, t_2]$. There is a local maximum for time t at point $t^*$, if there exists some $\varepsilon > 0$ such that $f(t^*) \geq f(t)$ when $|t - t^*| < \varepsilon$. Similarly, we count the local minima at $t^*$, if $f(t^*) \leq f(t)$ when $|t - t^*| < \varepsilon$. Creativity for actor i over time window ws is therefore: $RL_i = \#local\ minima_i^{ws} + \#local\ maxima_i^{ws}$

---

[1] Formerly *TeCFlow*.

*Patent applications*

Prior research has widely used patents as indicators of firms' ability to produce innovation (e.g. Guler and Nerkar, 2012; Lahiri, 2010; Fleming et al., 2007). Patents are granted only if they describe an invention that exceeds a minimum threshold of innovation: suitable for industrial application, nonobvious and novel. We use patent applications as our measure for innovation. For example, van Stel et al. (2014) found that for employer entrepreneurs in the European Union the number of patents applications is a predictor of long-term survival. Applications, rather than patents granted, is a more meaningful performance indicator for this study, because we are investigating the immediate effects in time between communication among firms and their innovative output. The time from when a patent application has been filed to the time the patent has been granted can be multiple years, a lag a firm can only partially influence. We collected communication data in 2005-2006, and thus collected patent applications in 2006 and 2007. We posit that communication in 2006 will have an influence on patent application production up to two years later.

Our choice of patents to represent innovation also relies on the fact that patents are among the few available variables we can collect and analyze to assess firms' performance. As most of the companies in the cluster are privately owned, other information – such as market capitalization, R&D expenditure and profits – are difficult to collect and often not publicly available. Note that we are measuring innovative performance of two types of companies (a) the start-ups, and (b) the Cambridge-based R&D organizations of large multinational biotech and pharmaceutical companies. For large multinational companies we only consider those patent applications where at least one inventor is from Massachusetts, based on the premise that "a start-up is a culture not delineated by metrics, and that a start-up can remain so at all ages and sizes" (Robehmed, 2013).

*Control variables*

We control for firm-size effects, and there is often an influence of firm age and size on the number of patent applications (Cohen and Levin, 1989; Scherer, 1965). *Age* is measured in years since the date of establishment. *Size* is measured as the number of employees working for the company – excluding those working in outside-cluster branches of large multinational corporations. Other parameters frequently used in analysis, such as R&D spending, market capitalization, profits, etc. were not available, as these companies are privately owned. The homogeneity of our sample – patent applications are all for new chemical compounds found or synthesized as a product of the research whose communication we analyzed – reduces the risk of exogenous effects and thus the need for additional control variables. The firms are all from the same industry – biotech, thus the data is free of the influence of other factors that might differ from industry to industry, and we do not need additional variables to control for these effects.

We also counted the number of communication ties connecting each biotech firm with universities and big pharmaceutical companies and assessed whether or not each biotech start-up communicated with non-profit research organizations and hospitals, creating a dichotomous variable. Hospitals and non-profit research organizations are represented as two aggregated nodes in our social network model, grouping in two nodes all institutions belonging to either the hospital or non-profit category.

## 4. Results

The communication network – illustrated in Figure 3 – clearly presents a core/periphery structure, with firms in the core representing a cohesive subgroup; the problem of measuring if, and how much, a firm is embedded in the network core is addressed using the core/periphery model introduced in (Borgatti and Everett, 1999).

**Figure 3.** Social network of the Boston Biotech cluster, size is degree centrality of node

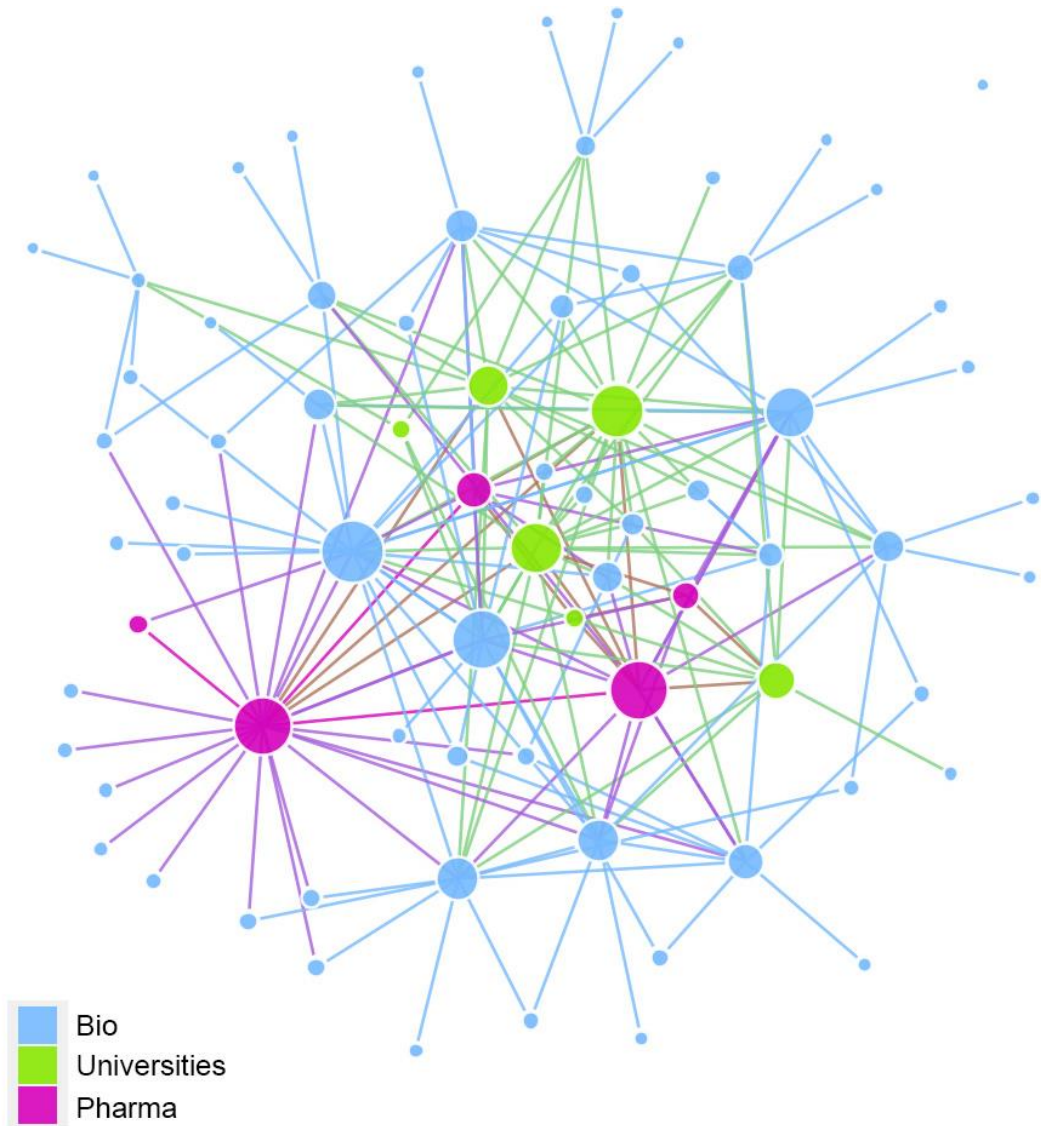

In Table 1, we provide the descriptive statistics for the variables presented in the previous section. The large employee numbers come from including the big pharma companies besides the startups.

**Table 1**. Descriptive statistics of the measured variables.

| Variable | M | SD | Min | Max |
| --- | --- | --- | --- | --- |
| Age | 9.97 | 6.72 | 1 | 27 |
| Size | 197.26 | 486.72 | 2 | 3017 |
| Indegree | 3.51 | 5.16 | 0 | 35 |
| Oscillation in Betweenness | 13.76 | 21.24 | 0 | 69 |
| Coreness | .06 | .09 | 0 | .49 |
| Distance from Cluster Center | 8.90 | 34.74 | 0 | 294.84 |
| Ties to Universities | 1.24 | 1.86 | 0 | 6 |
| Ties to Big Pharmas | .84 | 1.24 | 0 | 5 |
| Patents Applications 2006 | 4.04 | 7.55 | 0 | 31 |
| Patents Applications 2007 | 4.23 | 7.59 | 0 | 29 |

Figure 4 summarizes our key results (shown in Table 2). We find that innovative performance, measured through the number of patent applications, is a function of network position. Geographical proximity does not play any significant role. While the 15 firms in the Cambridge/Boston area have slightly fewer patent applications (28.73) than the 4 firms outside of Cambridge and Boston (30.75 patent applications), this difference is not statistically significant. The picture is very different however, for the 25 firms that are also located within the Cambridge/Boston area, but are not part of the communication core. These firms have only 7.44 patent applications, on average. The 27 firms outside of the Cambridge/Boston area, which are not part of the network core, perform the worst: they have (significantly) fewer patent applications, only 3.52 on average.

**Figure 4.** Summary of results (PA-Patent Applications).

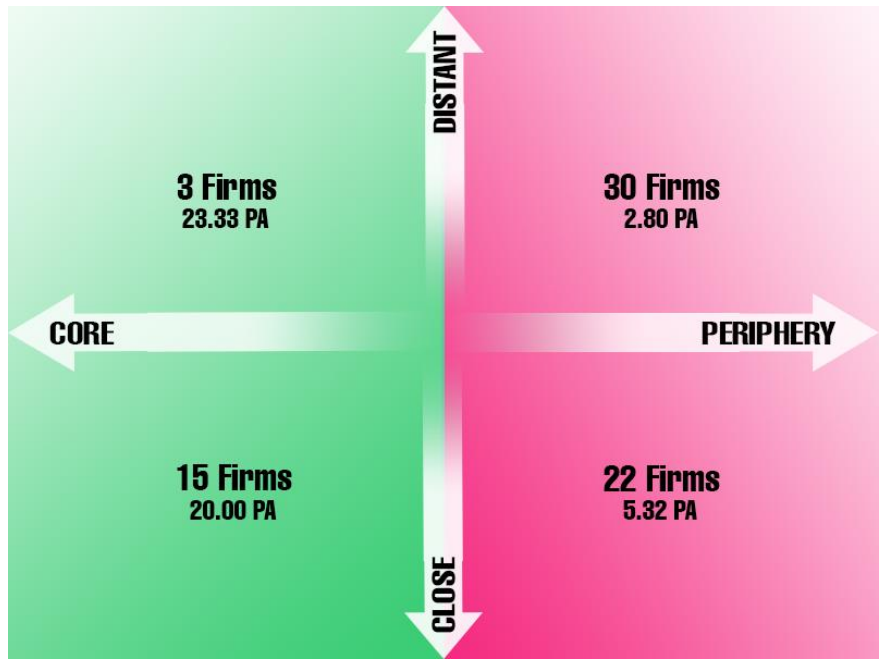

Results, in this first step of analysis, confirm the positive association of network coreness with patent applications from 2006 to 2007. There is no positive influence of geographical location alone on innovative performance.

**Table 2.** One-way ANOVA of patent applications in years 2006-2007 and post-hoc tests (Tukey HSD)

|  | Sum of Squares | df | Mean Square | F | Sig. | Mean | Post hoc analysis (Tukey HSD) | | | |
| --- | --- | --- | --- | --- | --- | --- | --- | --- | --- | --- |
|  |  |  |  |  |  |  | G1 | G2 | G3 | G4 |
| Between groups | 3833.03 | 3 | 1277.68 | 7.47 | .000 | G1 = 20.00 |  | *** | *** |  |
| Within groups | 11296.24 | 6( | 171.16 |  |  | G2 = 5.32 | *** |  |  |  |
| Total | 15129.27 | 6! |  |  |  | G3 = 2.80 | *** |  |  | * |
|  |  |  |  |  |  | G4 = 23.33 |  |  | * |  |

*Note.* *p < .1; **p < .05; ***p < .01.

G1= Firms in the core of the communication network, that are in a strong geographical proximity;

G2 = Firms at the periphery of the communication network, that are in a strong geographical proximity;

G3= Firms at the periphery of the communication network, that are in a weaker geographical proximity;

G4= Firms in the core of the communication network, that are in less geographical proximity (by postal area codes).

Table 3 reports the Pearson's correlation coefficients for the variables considered in our study. We basically find a strong positive correlation of patent applications with all our variables except with distance from the cluster center, suggesting that indeed, location does not matter. This is quite surprising, as it was found in a previous analysis of the same data, that location does matter, but only for the amount of communication between firms (Allen et al., 2009). It therefore seems that there must be a big difference between quantity and quality of networking.

**Table 3.** Pearson's correlation matrix of the measured variables.

|    | Variable | 1 | 2 | 3 | 4 | 5 | 6 | 7 | 8 | 9 | 10 |
|----|----------|---|---|---|---|---|---|---|---|---|----|
| 1  | Age | 1.00 | | | | | | | | | |
| 2  | Size | .44** | 1.00 | | | | | | | | |
| 3  | Indegree | .26** | .38** | 1.00 | | | | | | | |
| 4  | Betweenness Oscillation | .22 | .43** | .39** | 1.00 | | | | | | |
| 5  | Coreness | .24* | .59** | .51** | .70** | 1.00 | | | | | |
| 6  | Distance from Cluster Center | .05 | -.07 | -.09 | -.12 | -.11 | 1.00 | | | | |
| 7  | Ties to Universities | .11 | .32* | .18 | .88** | .65** | -.12 | 1.00 | | | |
| 8  | Ties to Big Pharmas | .15 | .62** | .51** | .80** | .87** | -.13 | .73** | 1.00 | | |
| 9  | Patents Applications 2006 | .48** | .61** | .52** | .48** | .56** | -.10 | .30* | .60** | 1.00 | |
| 10 | Patents Applications 2007 | .48** | .58** | .68** | .54** | .56** | -.11 | .32** | .56** | .93** | 1.00 |

Table 3 also shows a positive association of the firms' age with size, reflecting the fact that most firms become bigger as they age. In addition, patent applications are positively correlated with the firms' age and size, suggesting that the bigger and older firms are, the more likely they will apply for more patents. We find that indegree correlates the strongest with patent applications, suggesting that more popular startups, i.e. those which are reported as communication partners by more firms, produce more patent applications. We also find that betweenness oscillation correlates with patent applications, suggesting that rotating leadership, i.e. taking turns in network centrality over a one year period, which we have found in earlier work to be a proxy of creativity, is an indicator of more patent applications. In addition, more ties to universities and to big pharmaceutical companies correlate with more patent applications and a position closer to the network core, with the strongest effect for ties with the largest pharmaceutical companies. Significant correlations of the number of ties to universities and big pharmaceutical companies with betweenness oscillation show that more diverse ties are associated with a more dynamic communication style. To put it in other words, researchers at startups who oscillate more in their social network position, and talk more with universities and big pharmaceutical companies generate more patent applications.

**Table 4.** Predicting patents applications: Regression results

|  | Dependent: Patent Applications 2006 | | | Dependent: Patent Applications 2007 | | |
| --- | --- | --- | --- | --- | --- | --- |
| Variable | Model 1 | Model 2 | Model 3 | Model 4 | Model 5 | Model 6 |
| Age | .243* | .250* | .198* | .281* | .291** | .204* |
| Size | .008** | .005** | .005** | .007** | .004* | .003** |
| Coreness |  | 23.659** |  |  | 26.825** |  |
| Distance from Cluster Center |  | -.009 |  |  | -0.13 |  |
| Oscillation in Betweenness |  |  | .077* |  |  | .089** |
| Indegree |  |  | .285* |  |  | .574** |
| Constant | -.167 | -1.163 | -1.336 | -.184 | -1.299 | -1.978* |
|  |  |  |  |  |  |  |
| Adj R-Squared | .443 | .499 | .541 | .395 | .468 | .651 |
| N | 68 | 68 | 68 | 68 | 68 | 68 |

\* $p < .05$, \*\* $p < .01$.

We constructed a total of 6 regression models. As the regressions illustrate, the results are consistent for patent applications in both years 2006 and 2007, further proof of the robustness of our hypotheses. We assumed that using patent applications beyond 2007 was not appropriate, because of the time lag from when the communication data was collected (2005-2006). Models 1 and 4 – as expected – illustrate the importance of firm size and age for the number of patent applications. We use firm age and size as control in the other models. Results of Models 2 and 5 again support H1a and H1b: the closer the firms are to the network core (and not the geographic center), the more they apply for new patents. There is economic advantage for a firm to be in the communication core, while there is no economic advantage for a firm to be close to the geographical center of the cluster.

Regarding networking quality, in models 3 and 6, we find that indegree as a proxy for popularity, and oscillation in betweenness centrality as a proxy for creativity, explain innovativeness in the subsequent two years. We thus confirm the explanatory value of communication structure for innovative performance measured through patent applications, supporting hypotheses H2 and H3. Moreover, patent applications are the prelude to patents granted, which often bring strategic and economic advantages to the firm (Bloom and Van Reenen, 2002); it therefore seems that being embedded in the communication core of the cluster, and having more high quality collaboration ties with peers, universities and big pharmaceutical companies is advantageous for high-tech firms.

## 5. Discussion: more than communication?

Since the groundbreaking work of Allen (1984), the positive influence of co-location on communication is well known. People within easy walking distance talk exponentially more to each other. In earlier work using the dataset described in this paper, it was shown that the Allen curve also applies to companies, with communication between company researchers beyond a six-kilometer radius dropping off exponentially (Allen et al., 2009). The question of whether the increase in communication within the six-kilometer circle is beneficial to innovation has not been answered until now. In this study, we find that there is no correlation between the location of biotech start-ups and the amount of innovation. Rather we find that what really matters for innovative capability is active and engaged communication between researchers, not their location. To put it in other words, the increase in communication by co-located companies does not necessarily lead to an increase in innovation. We speculate that chitchat and small talk happening when researchers run into each other at talks or over lunch does not necessarily lead to the exchange of knowledge necessary for increased innovation. In the language of weak and strong ties, these location-induced ties might be rather weak, focused on friendship and maintaining social contact, and do not necessarily help innovation. On the other hand, the weekly surveys we collected, the basis for this analysis, asked specifically about communication on a scientific topic, i.e. respondents were primed to think about knowledge exchange with peers, potentially indicating stronger ties.

Compared to earlier work that has shown a connection between location, network, and firm performance (Decarolis and Deeds, 1999; Owen-Smith and Powell, 2004) we offer a more differentiated view. While geography indeed facilitates communication, it is not necessarily of the type leading to more innovation. There are now many examples where non-co-located knowledge workers have collaborated intensively over long-distance, using the Internet, without ever having met face-to-face (Van Noorden, 2014). What seems to matters is communication: the more biotech start-ups are part of the communication core, the more innovative they are, independent of their location.

On the theoretical level our research contributes to extant literature on the influence of location on innovation. We find a dichotomy between location and communication intensity on the one hand, and innovation and communication intensity on the other hand. While we find a potentially positive influence of location on social capital growth, the relationship between location and innovation is mitigated through communication intensity and quality – demonstrated through rotating leadership, i.e. changes in network position of researchers over time. Our research emphasizes the primacy of social capital over communication intensity. More research is needed to further investigate what influences the creation of social capital between firms.

The managerial implications of our research are twofold: firstly, location matters, but only if R&D staff members do not have the social network. It therefore makes sense for managers to invest into expensive real-estate near the biotech cluster center, to increase serendipitous interaction between researchers from different companies and foster knowledge spillover. Secondly, however, expensive real-estate investment can be replaced by other ways of maintaining social ties, for example, deployment of social networking software to encourage and enhance discussion, collaboration and engagement, along with an appropriate change towards more collaborative behavior. A more central position in the knowledge network is associated with a more intense communication of R&D members at startups with universities and large pharma companies; and we proved that a central position is beneficial to innovation. Accordingly, is recommended for managers to encourage members of their R&D departments to maintain informal social ties to their former research groups at the universities, as well as establish relationships with R&D departments at large pharmaceutical companies.

## 6. Limitations

Due to our firms being privately held, we have not been able to collect variables commonly used to control for differences in R&D spending, market capitalization, profits, etc. among the firms in our sample. However, the homogeneity of our sample addresses concerns about endogeneity as all our startups are pharma-biotech companies. Because all patent applications are filed for chemical compounds and processes by firms all active in the same subsector, we assume little difference in the propensity to file patent applications.

There are also limitations regarding the generalizability of our results, as our sample is restricted to one industry – biotech, and one area, the Boston region. However, as our data is homogeneous with respect to the industry, we posit that results can be generalized for other subsectors in the high-tech industry. Also, the greater Boston area is one of the world's leading regions of innovation, thus being representative of other high-tech innovation clusters (Porter et al., 2005).

## 7. Conclusion – it's all about quality

Marissa Mayer, the CEO of Yahoo, who has asked her employees to return to their office desks instead of telecommuting (Cain Miller and Rampell, 2013), might have gotten it wrong. Co-locating knowledge workers might get the dialog going, but may not necessarily induce innovation. Rather, it's a culture of knowledge sharing and rotating leadership that needs to be established (Orlikowski, 1993) independent of where the knowledge workers are located. Firms that are geographically distributed will want to examine their communication patterns and the relationship of those patterns to outcomes that are important to the firm to determine whether there is a culture of knowledge-sharing as well as the infrastructure and practices to support that culture.

**References**

Allen, T.J. (1984), *Managing the flow of technology: Technology transfer and the dissemination of technological information within the R&D organization*, MIT Press, Cambridge, MA.

Allen, T.J., Raz, O., and Gloor, P. (2009), "Does Geographic Clustering Still Benefit High Tech New Ventures?"*,* Working paper, Massachusetts Institute of Technology Engineering System Division, Cambridge, MA.

Aral, S. and Van Alstyne, M. (2007), "Network Structure & Information Advantage", Paper presented at the Academy of Management Conference, 3-8 August, Philadelphia, PA.

Barsky, N.P. (1999), "A Core/Periphery Structure in a Corporate Budgeting Process", *Connections*, Vol. 22 No. 2, pp. 1–29.

Battistoni, E. and Fronzetti Colladon, A. (2014), "Personality Correlates of Key Roles in Informal Advice Networks", *Learning and Individual Differences,* Vol. 34, pp. 63–69.

Bloom, N. and Van Reenen, J. (2002), "Patents, Real Options and Firm Performance", *The Economic Journal,* Vol. 112 No. 478, pp. C97–C116.

Borgatti, S.P. and Everett, M.G. (1999), "Models of core/periphery structures", *Social Networks,* Vol. 21 No. 4, pp. 375–395.

Boschma, R. (2005) "Proximity and Innovation: A Critical Assessment", *Regional Studies*, Vol.39 No.1, pp. 61-74.

Cain Miller, C. and Rampell, C. (2013), "Yahoo Orders Home Workers Back to the Office", *The New York Times*, 25 February.

Capaldo, A. and Petruzzelli, A. M. (2014), "Partner Geographic and Organizational Proximity and the Innovative Performance of Knowledge-Creating Alliances", *European Management Review*, Vol. 11 No. 1, pp. 63–84.

Cattani, G. and Ferriani, S. (2008), "A Core/Periphery Perspective on Individual Creative Performance: Social Networks and Cinematic Achievements in the Hollywood Film Industry", *Organization Science,* Vol. 19 No. 6, pp. 824–844.

Chuy, M., Zhang, J., Resendes, M., Scardamalia, M. and Bereiter, C. (2011), "Does contributing to a Knowledge Building dialogue lead to individual advancement of knowledge?", in *CSCL2011 Computer Supported Collaborative Learning: Connecting Research to Policy and Practice Conference Proceedings*, International Society of the Learning Sciences, Hong Kong, China, Vol. I, pp. 57-63.

Cohen, W.M. and Levin, R.C. (1989), "Empirical studies of innovation and market structure", in Schmalensee, R. and Willig, R.D. (Ed.), *Handbook of industrial organization 2*, Elsevier, New York, NY.

Cross, R. and Cummings, J.N. (2004), "Tie network correlates of individual performance in knowledge-intensive work", *Academy of Management Journal*, Vol. 47 No. 6, pp. 928–937.

Cummings, J. N. (2007), "Leading groups from a distance: How to mitigate consequences of geographic dispersion", in Weisband, S. (Ed.), *Leadership at a Distance: Research in Technology- Supported Work*, Lawrence Erlbaum Associates: Taylor and Francis Group, New York, NY.

Davis, J.P. and Eisenhardt, K.M. (2011), "Rotating leadership and collaborative innovation recombination processes in symbiotic relationships", *Administrative Science Quarterly*, Vol. 56 No. 2, pp. 159-201.

Decarolis, D.M. and Deeds, D.L. (1999), "The Impact of Stocks and Flows of Organizational Knowledge on Firm Performance: An Empirical Investigation of the Biotechnology Industry", *Strategic Management Journal,* Vol. 20 No. 10, pp. 953–968.

Demsetz, H. (1988), "The Theory of the Firm Revisited", *Journal of Law, Economics, and Organization,* Vol. 4 No. 1, pp. 141–161.

Festinger, L., Back, K.W. and Schachter, S. (1950), *Social pressures in informal groups; a study of human factors in housing*, Stanford University Press, Stanford, CA.

Fliaster, A. and Schloderer, F. (2010), "Dyadic ties among employees: Empirical analysis of creative performance and efficiency", *Human Relations*, Vol. *63* No. 10, pp. 1513-1540.

Fleming, L., Chen, D. and Mingo, S. (2007), "Collaborative Brokerage, Generative Creativity, and Creative Success", *Administrative Science Quarterly,* Vol. 52 No. 3, pp. 443–475.

Freeman, L. (1977), "A set of measures of centrality based on betweenness", *Sociometry*, Vol. 40, pp. 35–41.

Funk, R.J. (2013), "Making the Most of Where You Are: Geography, Networks, and Innovation in Organizations", *Accademy of Management Journal,* Vol. 57 No. 1, pp. 193–222.

Gloor, P.A., Kidane, Y.H., Grippa, F., Marmier, P. and Von Arb, C. (2008), "Location matters measuring the efficiency of business social networking", *International Journal of Foresight and Innovation Policy,* Vol. 4 No. 3-4, pp. 230–245.

Gloor, P. and Zhao, Y. (2004), "TeCFlow – A Temporal Communication Flow Visualizer for Social Network Analysis", paper presented at *ACM CSCW Workshop on Social Networks*, 6 November, Chicago, IL.

Gloor, P.A., Almozlino, A., Inbar, O., Lo, W. and Provost, S. (2014), "Measuring Team Creativity Through Longitudinal Social Signals", Unpublished Manuscript, *arXiv preprint arXiv:1407.0440*.

Goldman, S.R. and Scardamalia, M. (2013), "Managing, understanding, applying, and creating knowledge in the information age: Next-generation challenges and opportunities", *Cognition and Instruction*, Vol. 31 No. 2, pp. 255-269.

Granovetter, M.S. (1973), "The Strength of Weak Ties", *American Journal of Sociology*, Vol. 78 No. 6, pp. 1360–1380.

Grant, R.M. (1996), "Toward a knowledge-based theory of the firm", *Strategic Management Journal*, Vol. 17, pp. 109–122.

Grippa, F. and Gloor, P.A. (2009), "You are who remembers you. Detecting leadership through accuracy of recall", *Social Networks*, Vol. 31 No. 4, pp. 255-261.

Guler, I. and Nerkar, A. (2012), "The impact of global and local cohesion on innovation in the pharmaceutical industry", *Strategic Management Journal,* Vol. 33 No. 5, pp. 535–549.

Gumusluoglu, L. and Ilsev, A. (2009), "Transformational leadership, creativity, and organizational innovation", *Journal of Business Research*, Vol. 62 No. 4, pp. 461-473.

Kidane, Y. and Gloor, P. (2007), "Correlating temporal communication patterns of the Eclipse open source community with performance and creativity", *Computational & Mathematical Organization Theory*, Vol. 13 No. 1, pp. 17–27.

Lahiri, N. (2010), "Geographic Distribution of R&D Activity: How Does it Affect Innovation Quality?", *Academy of Management Journal,* Vol. 53 No. 5, pp. 1194–1209.

Massachusetts Biotechnology Council (2014), *Impact 2020. Advancing Massachusetts Leadership in the Life Sciences For Patients*, Cambridge, MA.

Nonaka, I. (1994), "A Dynamic Theory of Organizational Knowledge Creation", *Organization Science,* Vol. 5 No. 1, pp. 14–37.

Orlikowski, W. (1993), "Learning from notes: Organizational issues in groupware implementation", *The Information Society,* Vol. 9 No. 3, pp. 237–250.

Owen-Smith, J. and Powell, W.W. (2004), "Knowledge Networks as Channels and Conduits: The Effects of Spillovers in the Boston Biotechnology Community", *Organization Science*, Vol. 15 No.1, pp. 5-21.

Porter, K., Whittington, K.B. and Powell, W.W. (2005), "The Institutional Embeddedness of High-Tech Regions: Relational Foundations of the Boston Biotechnology Community", in Breschi, S. and Malerba, F. (Ed.), *Clusters, Networks, and Innovation*, Oxford University Press, Oxford, U.K.

Pouder, R. and St. John, C.H. (1996), "Hot spots and blind spots: geographical clusters of firms and innovation", *Academy of Management Review,* Vol. 21 No. 4, pp. 1192–1225.

Raz, O. and Gloor, P.A. (2007), "Size Really Matters—New Insights for Start-ups' Survival", *Management Science*, Vol. 53 No. 2, pp. 169–177.

Robehmed, N. (2013), "What Is A Start-up?", *Forbes*, 16 December.

Rotolo, D. and Petruzzelli, A.M. (2013), "When does centrality matter? Scientific productivity and the moderating role of research specialization and cross-community ties", *Journal of Organizational Behavior*, Vol. 34 No. 5, pp. 648–670.

Saxenian, A. (1994), *Regional Advantage: Culture and Competition in Silicon Valley and Route 128*, Harvard University Press, Cambridge, MA.

Saxenian, A. (2006), *The New Argonauts: Regional Advantage in a Global Economy*, Harvard University Press, Cambridge, MA.

Scherer, F.M. (1965), "Firm size, market structure, opportunity, and the output of patented inventions", *The American Economic Review,* Vol. 55 No. 5, pp. 1097–1125.

Sinnott, R.W. (1984), "Virtues of the Haversine", *Sky Telescope,* Vol. 68 No. 2, pp. 159.

Sparrowe, R.T., Liden, R.C., Wayne, S.J. and Kraimer, M.L. (2001), "Social networks and the performance of individuals and groups", *Academy of Management Journal,* Vol. 44 No. 2, pp. 316–325.

Teece, D.J. (1986), "Profiting from technological innovation: Implications for integration, collaboration, licensing and public policy", *Research Policy,* Vol. 15, pp. 285–305.

Uzzi, B. (1996), "The Sources and Consequences of Embeddedness for the Economic Performance of Organizations: The Network Effect", *American Sociological Review,* Vol. 61 No. 4, pp. 674–698.

Uzzi, B. and Spiro, J. (2005), "Collaboration and Creativity: The Small World Problem", *American Journal of Sociology,* Vol. 111 No. 2, pp. 447–504.

Van Noorden, R. (2014), "Online collaboration: Scientists and the social network", *Nature*, Vol. 512, pp. 126–129.

Van Stel, A., Millán, J.M. and Román, C. (2014), "Investigating the impact of the technological environment on survival chances of employer entrepreneurs", *Small Business Economics*, Vol. 43 No. 4, pp. 839–855.
25

Woolley, A.W., Chabris, C.F., Pentland, A., Hashmi, N. and Malone, T.W. (2010), "Evidence for a collective intelligence factor in the performance of human groups", *Science*, Vol. 330, pp. 686-688.

Zaheer, A. and Bell, G.G. (2005), "Benefiting from Network Position: Firm Capabilities, Structural Holes, and Performance", *Strategic Management Journal,* Vol. 26 No. 9, pp. 809–825.